\documentclass[letterpaper]{article}
\usepackage{aaai}
\usepackage{times}
\usepackage{helvet}
\usepackage{courier}
\usepackage{graphicx}
\usepackage{caption}
\usepackage{subcaption}
\usepackage{url}
\nocopyright

\begin{document}
%
\title{Crowdsourcing Information Extraction for Biomedical Systematic Reviews}
\author{\large Yalin Sun, %
       Pengxiang Cheng, %
       Shengwei Wang, %
       Hao Lyu, %
       Matthew Lease, %
       Iain Marshall\textsuperscript{1}, \and %
       Byron C.\ Wallace\textsuperscript{2} \\
\\
University of Texas at Austin~~~~~ %
\textsuperscript{1} King's College London~~~~~ %
\textsuperscript{2} Northeastern University %
    }
\maketitle

\section{Abstract}
Information extraction is a critical step in the practice of conducting biomedical systematic literature reviews. Extracted structured data can be aggregated via methods such as statistical meta-analysis. Typically highly trained domain experts extract data for systematic reviews. The high expense of conducting biomedical systematic reviews has motivated researchers to explore lower cost methods that achieve similar rigor without compromising quality. Crowdsourcing represents one such promising approach. In this work-in-progress study, we designed a crowdsourcing task for biomedical information extraction. We briefly report the iterative design process and the results of two pilot testings. We found that giving more concrete examples in the task instruction can help workers better understand the task, especially for concepts that are abstract and confusing. We found a few workers completed most of the work, and our payment level appeared more attractive to workers from low-income countries. In the future, we will further evaluate our results with reference to gold standard extractions, thus assessing the feasibility of tasking crowd workers with extracting biomedical intervention information for systematic reviews.

\section{Introduction}

Systematic literature review for clinical trials is a well-established research method in the medical field, underpinning \emph{evidence-based medicine} (EBM) \cite{sackett}. Systematic reviews influence healthcare practice, health-related policy making, and academic research \cite{chang2004interventions}. However, the process of conducting a systematic reviews is laborious, time-consuming, and expensive. Typically, (highly) trained domain experts are recruited to extract targeted information to be combined via statistical meta-analysis. Such reviews are critical components of EBM, but existing methods do not scale to the rapidly growing evidence base: new approaches are thus needed to expedite the process \cite{wallace2013modernizing}.

Recently efforts have been made within the natural language processing (NLP) community to automatically extract biomedical information from texts \cite{jonnalagadda2015automating}. However, accuracy using machine extraction alone is underwhelming \cite{dumitracheachieving}. Crowdsourcing has previously been applied to information extraction (IE) tasks \cite{kondreddi2014combining,burger2014hybrid}, but no prior work has investigated the feasibility of using the crowd for biomedical data extraction to support systematic reviews.

In this study we explore the feasibility using crowdworkers to extract data for systematic reviews. Specifically, we designed a task asking crowd workers to extract structured data pertaining to treatments administered in clinical trials described in the abstracts of medical publications. Results of this experiment will benefit future research in NLP, biomedical systematic review, and crowdsourcing by examining the feasibility and quality of using crowd workers to perform this type of specialized task at scale. This project is in progress, thus we only report the task design and results of the pilot tests that we have conducted with crowd workers.  

\section{Task Requirements \& Design}

\begin{figure}[!htb]
\centering
\includegraphics[width=1\linewidth]{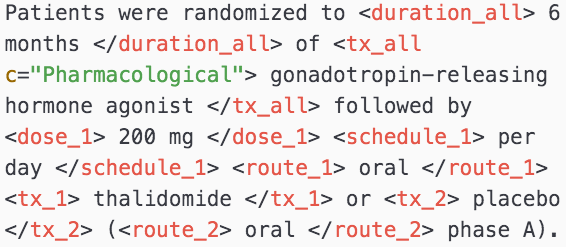}
\caption{Information Extraction Example}
\label{fig::annotation}
\end{figure}

Our full dataset contains 96 abstracts of articles describing randomized controlled trials (RCTs). We used CrowdFlower as the experiment platform. The goal of this crowdsourcing task is for human workers to identify and extract intervention (i.e., treatment) information such as duration, schedule, dose, and route from those abstracts (e.g., Figure \ref{fig::annotation}).  

One of the major challenges in designing a crowdsourcing task for complex extraction work is to make the information extracted by non-professional workers as structured as possible. Our first attempt was to design an interactive interface for workers to add annotations directly to an abstract text. But we found that the platform we chose did not support this. Therefore, we instead asked workers to manually enter (copy/paste) extracted information into distinct text boxes designated for each field.  

In CrowdFlower, each abstract record was regarded as a row and the information extracted from a worker about a specific abstract was considered a judgment. CrowdFlower organizes tasks by `pages'; several abstracts can be placed into one page. We decided to collect three judgments for each abstract and present 5 abstracts at a time on each page. We pay workers \$0.30 for the completion of each page. As a quality control mechanism, we selected 23 relatively `easy' abstracts to serve as test examples. CrowdFlower randomly selects several of these as a pre-screening quiz before workers can begin work. The remaining are randomly inserted into each pages as `honeypots' to encourage consistency.


Furthermore, to minimize individual differences and enhance reliability, we recruited three individuals working on the same abstract and aggregated their annotations via majority voting. All of the above design elements were tested in two pilot testings and adjusted accordingly according to the pilot results. In the next section, we will describe the results of the pilot testings and the iterative revisions of the task design.

\section{Pilot Tests}

To evaluate the effectiveness of our design, we launched two rounds of pilot testings on CrowdFlower in November 2015. In the pilot tests we kept all the settings mentioned above except the number of abstracts being used. We only included a subset of our data (20 unlabeled abstracts). In total, 60 judgments were collected (20 abstracts x 3 judgments/abstract).  

In pilot testing I, all 60 judgments were finished in 5 hours with a cost of \$5. Twenty workers started the task, but only six of them passed the pre-qualification tests. According to the results and analyses in the first pilot testing, we made several modifications to our original task design.
\begin{itemize}

\item
Since workers left many fields empty, we added a ``required'' validator to every field shown. Further, we also emphasized in the instructions that workers should input NA for any fields that they thought have no relevant information existing in the abstract. 

\item
Some intervention information is shared among different arms in trials (i.e., intervention groups). To account for this complexity, we added some logic to the task design by adding a checkbox beside every field in the shared section.

\item
To help workers better understand the task, we wrote more detailed instructions addressing all the issues we discovered in the pilot testing. We also added more labeling examples, trying to cover more varieties on what information should be extracted.
\end{itemize}

In pilot testing II, we inherited the same task settings as of the first pilot testing and added the constraint that workers belong to English-speaking countries.

It took one day this time to complete data collection. In total, four workers provided the 60 judgments. Among the qualified workers, the accuracy rate of test questions reached 94.3\% and we found a significant boost in the accuracy rate in the second pilot testing. In addition, the ratings of our job in terms of overall satisfaction, instruction clarity, test question fairness, ease of job were all improved compared with those in the first pilot round. One exception is the evaluation of payment. This might because we only recruited workers from English-speaking countries in the second pilot. Review of the results from the second pilot testing revealed that the formats of the answers from the workers were much cleaner, and a higher consistency was found among the workers\footnote{A video about this project can be found at \url{www.ischool.utexas.edu/~claire/crowdsourcing.mp4}.}.

\section{Conclusions}
The preliminary results of our pilot testings show that it is promising to recruit workers from crowdsourcing platforms to conduct the task of data extraction for systematic reviews. This task is inherently complex, and intuitively we might expect laypersons to struggle with it. However, we found that with clear instructions and appropriate incentives, crowd workers can perform this type of task with reasonable accuracy, and at costs much lower than domain experts. Our pilot experiments also revealed an improvement of the overall quality with the adjustments in the task design.

There are some important limitations to our experiment. First, the specific crowdsourcing platform we selected (CrowdFlower) influenced our task design and may have potentially affected the quality of the data collected. Although it supported most of the workflow and interface requirements, there are still several weaknesses we identified (e.g., interface design limitations and lack of interaction support). In the future, we can consider designing and deploying this task via other platforms such as Amazon MTurk or Zooniverse to see if there is a more optimal solution.

Second, the compensation and number of annotations collected for each abstract were somewhat arbitrary; we followed the setting of previous similar tasks and our budget. This may affect the quality of the results we collected. For example, in the second pilot test, some workers noted that they felt the pay was unfair; this should probably be upwardly adjusted in future work. 

Preliminary results comparing the collected data with `gold standard' annotations are promising. Preliminary analysis suggests that for some fields -- including sample size, treatment group sizes and intervention names -- crow annotations correspond to reasonably high accuracy (\( >0.70 \)). Going forward, we will further analyze the dataset and examine the overall quality of the information extracted by the crowd workers and the feasibility of using crowdsourcing approach for biomedical intervention information extraction at a larger scale. We also aim to explore hybrid machine learning/crowdsourcing approaches that actively involve domain experts.

\section{Acknowledgments}
The authors would like to thank CrowdFlower for the free credits we received for this study and the crowd workers who participated in this research. Research reported in this article was supported in part by the NCI of the National Institutes of Health (NIH) under award number UH2CA203711.

\bibliographystyle{aaai} 

\bibliography{crowdsourcing_status_report}

\end{document}